\documentclass[final,5p,times,preprint]{elsarticle}
\usepackage[utf8]{inputenc}
\usepackage{amsfonts}
\usepackage{amsmath}
\usepackage{amssymb}
\usepackage{bm}
\usepackage{lineno,hyperref}
\usepackage{comment}
\usepackage{xcolor}
\usepackage{booktabs}
\modulolinenumbers[5]
\usepackage[normalem]{ulem}

\newcommand{\e}{\varepsilon}

\def\amm#1{\textcolor{purple}{#1}} 
\def\lay#1{\textcolor{blue}{#1}} 
\def\jgc#1{\textcolor{olive}{#1}} 

\journal{Physics Letters B}

\biboptions{comma,sort&compress}

\begin{document}

\begin{frontmatter}

\title{Determining $B(E1)$ distributions of weakly bound nuclei from breakup cross sections using 
Continuum Discretized Coupled Channels calculations. Application to $^{11}$Be}


\author[FAMN,iC1]{A. M. Moro\corref{mail}}
\cortext[mail]{Corresponding author}
\ead{moro@us.es}

\author[FAMN,iC1]{J. A. Lay}

\author[FAMN,CNA]{J. G\'omez Camacho}


\address[FAMN]{Departamento de F\'{\i}sica At\'omica, Molecular y Nuclear, Facultad de F\'{\i}sica, Universidad de Sevilla, Apartado 1065, E-41080 Sevilla, Spain}

\address[iC1]{Instituto Interuniversitario Carlos I de F\'isica Te\'orica y Computacional (iC1), Apdo.~1065, E-41080 Sevilla, Spain}

\address[CNA]{Centro Nacional de Aceleradores, U. Sevilla, J. Andalucía, CSIC, Avda Thomas A Edison, 7, E-41092 Sevilla, Spain }


\begin{abstract}
A novel method to extract the $B(E1)$ strength of a weakly bound nucleus  from experimental Coulomb dissociation data is proposed. The method makes use of  continuum discretized coupled channels (CDCC) calculations, in which both nuclear and Coulomb forces are taken into account to all orders. This is a crucial advantage with respect to the standard procedure based on the Equivalent Photon Method (EPM) which does not properly take into account nuclear distortion, higher order coupling effects, or Coulomb-nuclear interference terms. 
The procedure is applied to  the $^{11}$Be nucleus using two sets of available experimental data at different energies, for which seemingly incompatible $B(E1)$ have been reported using the EPM. We show that the present procedure gives consistent $B(E1)$ strengths, thus solving the aforementioned long-standing discrepancy between the two measurements. 
\end{abstract}


\end{frontmatter}


\section{Introduction}
\label{sec:intro}
The investigation of nuclei close to the neutron and proton driplines require measuring observables which display their unusual structure properties. 
A relevant question is how the electromagnetic field connects the ground state of a weakly bound nucleus to its continuum. For that, one   would ideally like to place the system under the action of a pure electromagnetic pulse, and observe the energy distribution of its fragments. In practice, this can be achieved experimentally by means of nuclear collisions, although these are sensitive not only to the Coulomb interaction but also to the nuclear interaction. 
By a suitable choice of the target, and an adequate range of scattering angles and collision energies, one can reduce the effect of the nuclear interaction, and have a Coulomb-dominated breakup reaction. Furthermore, under appropriate kinematical conditions, one can assume a simplified, 
first-order description of the reaction mechanism which leads to a proportionality of the observed experimental quantity, the breakup cross section distribution, with the structure property to be determined, which is the electric dipole $B(E1)$ distribution.  This is the Equivalent Photon Method (EPM), for which the double differential breakup cross section, as a function of the scattering angle $\theta$ and the breakup energy $\varepsilon$ is given by
\begin{equation}
 \label{EPM} 
    \frac{d^2 \sigma}{d \Omega d \e} = \frac{d B(E1, \varepsilon)} {d\varepsilon} F_1(\theta, \xi), 
\end{equation}
where $F_1(\theta,\xi)$ is the dipole Coulomb excitation function, which depends on the scattering angle and on the Coulomb adiabaticity parameter $\xi$, which is proportional to the breakup energy $\varepsilon$. 
This function was derived in the seminal work of Coulomb excitation of Alder and Winther \cite{AW75}. At high energies, relativistic effects must be taken into account. This can be done using the generalization of  Bertulani and Baur~\cite{Ber88}, in which the Coulomb excitation function is replaced by the number of virtual photons produced by the target  $N_{E1}( \theta, \xi)$. They are related as:
\begin{equation}
 F_1(\theta, \xi)= \frac{16\pi^3}{9\hbar c}\frac{dN_{E1}( \theta, \xi)}{d\Omega} . \label{EPM3} 
\end{equation}

Practical application of a Coulomb dissociation experiment involves considering a certain experimental angular range, determined by the experimental setup, over which the double differential cross section is integrated.  Also, the measurements are performed at certain nominal breakup energies $\e_i$, which incorporate a distribution of nearby energies. Thus, the measured quantities are a discrete set of averaged differential cross sections $\sigma_i$, which, within the EPM approach, are given by
\begin{equation}
\label{EPMINT}
 \sigma_i =  {\overline B}(E1, \e_i) {\overline F}_1(\e_i)  ,
\end{equation}
where ${\overline F}_1(\e_i)$ is the dipole Coulomb function integrated over the angular and excitation energy ranges. 
Note that the value $ {\overline B}(E1, \e_i)$ extracted from Eq.~(\ref{EPMINT}) should be understood as an average of the $B(E1)$ distribution over the energy range represented by $\e_i$, with weights determined by the integral of $F_1(\theta,\xi)$ over the angular range. 
This fact complicates the comparison of ${\overline B}(E1, \e_i)$ values obtained from different experiments, as well as these with theoretical calculations.

There are many approximations implicit in expressions (\ref{EPM}) and (\ref{EPMINT}). First, the semiclassical treatment should be valid, so that the scattering angle defines uniquely a classical trajectory, which is assumed to provide an accurate description of the quantum mechanical wave function. Second, the trajectories should be pure Coulomb, and should not be affected by the ever-present nuclear interaction. Third, the coupling interactions should be purely dipole Coulomb (no nuclear coupling), and have the asymptotic $r^{-2}$ dependence over all the relevant range. Fourth, a first-order perturbation treatment of the Coulomb dipole force should be valid. Fifth, the effect of higher multipoles on breakup cross sections should be neglected. Moreover, the application of the integrated expression  (\ref{EPMINT}) requires that the aforementioned approximations should be valid for all the scattering angles contained in the experimental angular range.

In actual experiments, it can be argued that the EPM approximation is ``fairly good'', assuming that heavy targets are used, small angles are measured and the collision energy is adequate. This regime is optimistically referred to as ``safe Coulomb'' (see, e.g.,~Refs.~\cite{Gla98,Gad03}). However, even in these ``safe Coulomb'' cases, the EPM may have non-negligible deviations from more accurate calculations
\cite{Esb95,Esb02,Pes17}, which would go as uncontrolled systematic uncertainties to the $B(E1)$ distributions  obtained from the breakup cross section using Eq.~(\ref{EPMINT}). Nuclear effects are sometimes taken into account by expressing the breakup cross sections $\sigma^e_i$ as a sum of a nuclear contribution $\sigma^n_i$ and a pure dipole Coulomb contribution, 
\begin{equation}
 \sigma^e_i = \sigma^n_i +  {\bar B}(E1, \e_i) {\bar F}_1(\e_i) \label{EPMexp} . 
\end{equation}
The former is obtained experimentally re-scaling cross sections on nuclear-dominated reactions  \cite{Pal03}. This procedure, however, neglects Coulomb-nuclear interference terms, as well as dynamical effects which may be very different in Coulomb and nuclear dominated reactions.  

Taking into account the enormous efforts devoted to perform such experiments, aimed at getting $B(E1)$ distributions with the highest possible accuracy, it is timely to understand the limitations of the EPM method and, whenever the approximations stated above are not well justified, 
substitute it by more accurate procedures based on the best quantum mechanical calculations available for the breakup cross sections. 
With this motivation, in this work we propose a new procedure to extract the $B(E1)$ distribution from Coulomb dissociation experiments, which relies on the Continuum-Discretized Coupled Channels (CDCC) method. CDCC is a well established fully quantum-mechanical reaction framework  which does not require
the approximations inherent to the EPM and overcomes most of its limitations. The procedure is applied to shed light on the apparently inconsistent $B(E1)$ distributions of $^{11}$Be extracted from  two different Coulomb dissociation experiments \cite{Fuk04,Pal03}.



\section{Theoretical procedure}
\label{sec:theo}
We start with a structure model calculation for the projectile, which is sufficiently amenable to be used as an input for a full quantum mechanical scattering calculation for the reaction process. For halo nuclei, a convenient choice is a few-body model, in which the projectile is described by a core and one or two valence particles, with the core being described by a small number of discrete states. The model will provide normalizable wavefunctions for the projectile ground state and a  set of continuum states, from which a $B(E1)$ distribution,  ${dB^0(E1,\e)}/{d\varepsilon}$, can be derived, which should be regarded  only as an initial estimate of the $B(E1)$ distribution to be extracted from experiment. 
The continuum states of the projectile can be discretized into a 
set of normalizable wavefunctions which, along with the Coulomb and nuclear potentials describing the interaction of the target with the fragments of the projectile,  can be taken as an input for a full quantum mechanical scattering calculation. 
In this work we adopt the extended Continuum-Discretized Coupled-Channels (XCDCC) method \cite{Sum06,Die14}. The calculation, suitably integrated over the experimental setup, and including the angular and energy resolution,  produces  model differential cross sections $\sigma^0_i$, evaluated at the experimental energies $\e_i$. The model cross sections can be compared with the experimentally measured cross sections $\sigma^e_i$. The results will not coincide in general, as it should be expected from the fact that the model $B(E1)$ distribution does not coincide with the actual $B(E1)$ distribution of the projectile. However, we can use the model as a tool to investigate the relation between the $B(E1)$ distribution and the breakup cross section, which will be much more accurate  than the EPM relation, Eq.~(\ref{EPMINT}), because it incorporates elements (quantum effects, nuclear forces, higher order coupling, etc) which are absent in the EPM.

As shown in the  \ref{apen:amplitudes}, we can introduce small changes in the model, by multiplying the different Coulomb dipole matrix elements by arbitrary factors close to one. 
%
This modifies the $B(E1)$ distribution at each measured energy $e_i$, 
\begin{equation}
B^m(E1,\e_i) \simeq  B^0(E1,\e_i)  (1+ 2 \delta(\e_i)).
 \label{neweqBdelta} 
 \end{equation}
 where $\delta(e_i)$ is an energy dependent factor defined in Eq.~(\ref{delta}). A remarkable result, Eq.~(\ref{sigmadelta}), is that the changes in the cross sections are  determined by the same quantities $\delta(e_i)$:
\begin{equation}
   \sigma^m_i \simeq   \sigma^0_i + \delta(\e_i) \,{\sigma^\prime}_i .
   \label{neweqSdelta}
\end{equation}

From Eqs.~(\ref{neweqBdelta}) and (\ref{neweqSdelta}), one can eliminate the explicit dependence in $\delta(\e_i)$, 
 leading to a relation between the $B(E1)$ distribution and the cross sections in the modified model. 
\begin{equation}
B^m(E1,\e_i) \simeq  B^0(E1,\e_i)   
  \left( 1 + 2 \frac{\sigma^m_i - \sigma^0_i}{\sigma^\prime_i} \right) .
 \label{quadraticnew} 
\end{equation}
This gives an approximate linear relationship between $B^m(E1,\e_i)$ and $\sigma^m_i$ which holds reasonably well, as it will be shown later in Fig.~\ref{fig:sigma_vs_be}.

The quantity ${\sigma'}_i$ is the key magnitude that encodes the relation of cross sections and $B(E1)$ values. It plays the role of the dipole excitation function ${\bar F}(\e_i)$ in the EPM, and can be obtained from model calculations following Eq.~(\ref{sigmap}) of the appendix. The actual $^{11}$Be system will, admittedly, be much more complex than the adopted model. However, it is reasonable to consider that a realistic description of  $^{11}$Be is compatible with the model calculation where the electric dipole matrix elements have been suitably adjusted.
So we are entitled to replace  the quantity $\sigma^m_i$ in Eq.~(\ref{quadraticnew}) by the measured value $\sigma^{e}_i$  and then infer an ``experimental'' value for the $B(E1)$ distribution as 
%
%
\begin{equation}
  B^{e}(E1,\e_i) = B^0(E1,\e_i)\left( 1 + 2 \frac{\sigma^{e}_i - \sigma^0_i}  {{\sigma^{\prime}}_i} \right) \label{Bexpnew} .
  \end{equation}
The obtained values are graphically depicted by the vertical lines in Fig.~\ref{fig:sigma_vs_be}, to be discussed later.

 It should be stressed that the $B(E1)$ values obtained by this procedure are unfolded from the experimental energy resolution, because the values of $\sigma_i^0$ and $\sigma^{\prime}_i$  are calculated integrating over the same energy and angular resolution of $\sigma_i^e$, taking  into account the energy dependence of the model $B(E1)$ distribution $d B^0(E1,\varepsilon)/d\varepsilon$.


\section{Application to $^{11}{\rm Be}$}
\label{sec:calculos}
We will apply the outlined procedure to the extraction of the $B(E1)$ distribution of $^{11}$Be. We consider two experiments carried out for this purpose using the reaction  $^{11}$Be on $^{208}$Pb. The first one is the experiment by Palit {\it et al.}~\cite{Pal03} performed at GSI at 
520~MeV/u. The other experiment was performed by Fukuda {\it et al.}~\cite{Fuk04} at RIKEN at 69~MeV/u. Both experiments measured breakup cross sections, and derived the $B(E1)$ distribution making use of the EPM, producing results that are not compatible, specially at low breakup energies. 
In this context, we note that a recent {\it ab-initio} calculation by Calci {\it et al} \cite{Cal16}, based on the no-core shell model with continuum (NCSMC), predicts a $B(E1)$ distribution in good agreement with the one extracted in the RIKEN experiment \cite{Fuk04}. However, a recent eikonal calculation performed in Ref.~\cite{Mos19} for the GSI data, using a structure model adjusted to reproduce the long-range features of the {\it ab-initio} calculation, overestimates the energy differential cross section from this experiment at the peak.


 In the present study, the  $^{11}$Be structure is 
 described using a two-body ($n$+$^{10}$Be) particle-plus-rotor model (PRM) with the Hamiltonian of Ref.~\cite{Sum07}.  
To account for the coupling with the $2^+$ state of the $^{10}$Be core, the $n$+$^{10}$Be central potential is deformed using a deformation parameter $\beta_2$ = 0.67,   
giving rise to core-excited admixtures in the $^{11}$Be states. 
In Ref.~\cite{Sum07}, several sets of parameters are considered for the central and spin-orbit parts, which result in different $B(E1)$ strengths. In this work, we present results with the sets III and V of Table I of \cite{Sum07} which will be denoted hereafter as S3 and S5, respectively. 

We have performed continuum-discretized coupled-channels calculations, including the $^{10}$Be excitation (XCDCC) \cite{Sum06,Die14}. These calculations  
 require the optical model potentials for $n$-$^{208}$Pb and  $^{10}$Be-$^{208}$Pb, with the latter including quadrupole deformation to account for possible excitations of $^{10}$Be during the reaction. For the reaction at 520~MeV/u, the $n$-$^{208}$Pb potential was generated by folding the Paris-Hamburg 
 $g$-matrix  NN effective interaction \cite{Ger83,Rik84} with the ground-state density of the target, obtained from a Hartree-Fock 
 calculation. For the reaction at 69~MeV/u, the $n$-$^{208}$Pb potential was taken from the global parameterization of Koning and Delaroche \cite{KD03}. The  $^{10}$Be-$^{208}$Pb potential  consists of a double folding of the projectile and target densities with an effective  $g$-matrix NN interaction, appropriate for each energy regime, namely, the  Br\`uyeres Jeukenne-Lejeune-Mahaux  \cite{Bau98,Bau01} for the 69~MeV/u data (see  also \cite{Pang-private,Xu13})  and the CEG07 interaction \cite{Furu-private,Fur12} for the 520~MeV/u data. 
Relativistic corrections were taken into account in both calculations, following \cite{Mor15}. The calculated differential cross sections were convoluted with experimental angular  and energy resolutions quoted in Refs.~\cite{Pal03,Fuk04}.

\begin{figure}[tb]
\begin{center}
 {\centering \resizebox*{0.80\columnwidth}{!}{\includegraphics{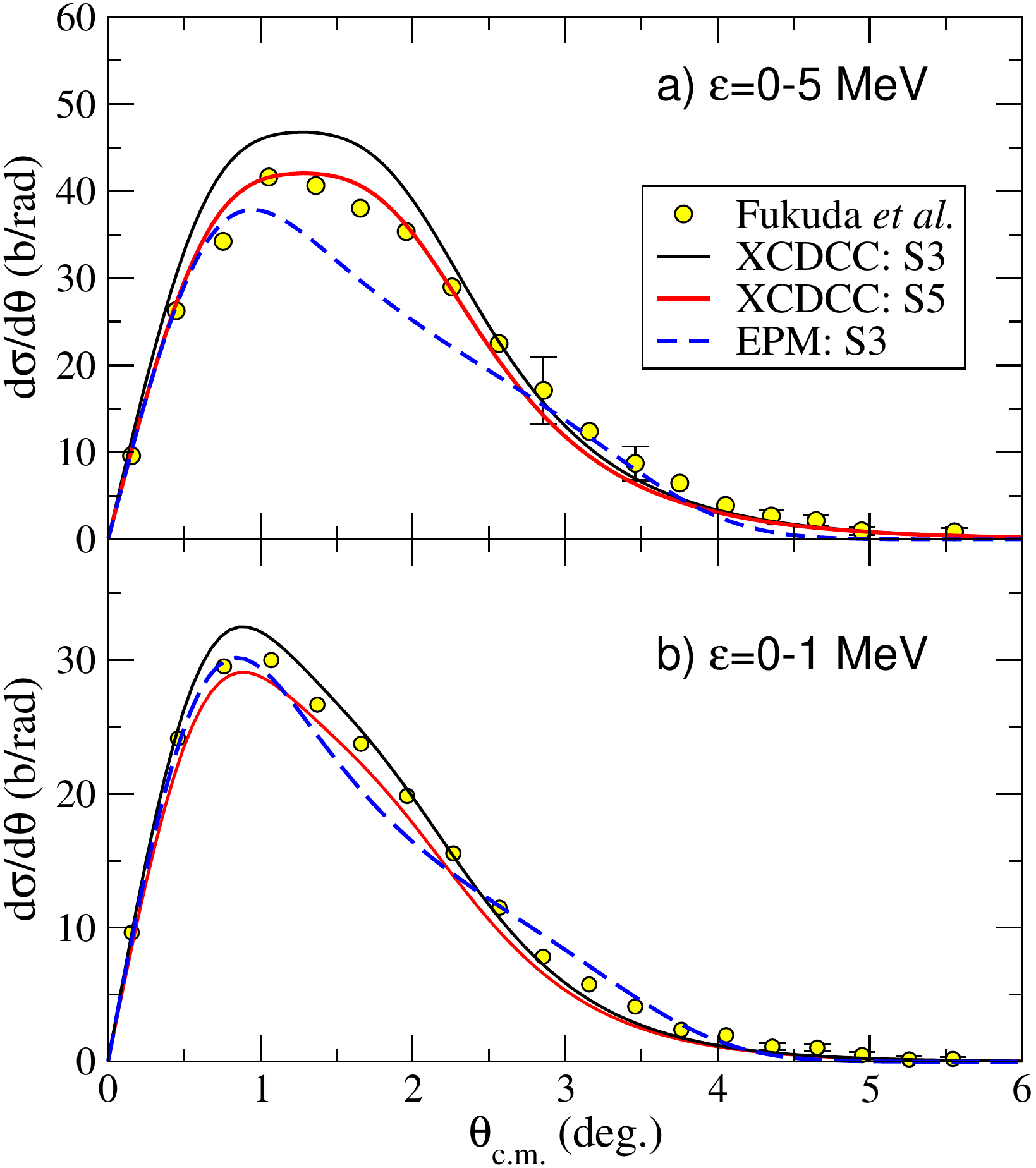}} \par}
\caption{\label{fig:angular69} Differential angular cross section for $^{11}$Be+$^{208}$Pb breakup at 69~MeV/u. The solid curves are XCDCC calculations with the S3 and S5 structure models, whereas the dashed line is the EPM result with model S3. All calculations have been convoluted with the experimental resolution~\cite{Fuk04}.
}
\end{center}
\end{figure}

 The experimental and calculated breakup angular distributions  for the incident energy at 69~MeV/u are shown in Fig.~\ref{fig:angular69}. It can be seen that the XCDCC method gives significantly larger cross sections as compared to the EPM calculations. Moreover, the XCDCC calculation reproduces well the shape of the angular distribution, even at relatively large scattering angles, for which the nuclear interaction will be relevant.  
 The 
 EPM angular distributions differ significantly from the data. 

\begin{figure}[tb]
\begin{center}
 {\centering \resizebox*{0.85\columnwidth}{!}{\includegraphics{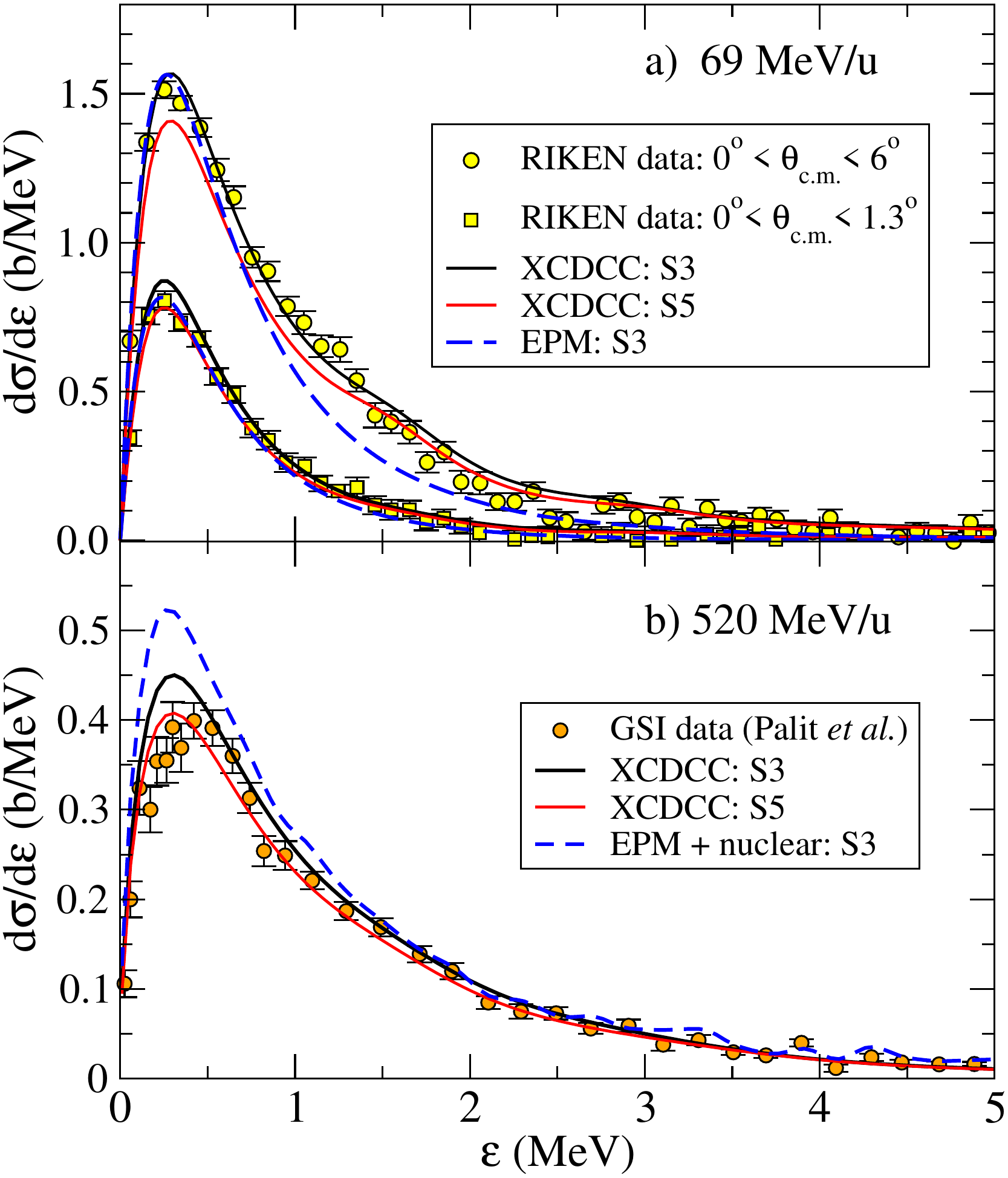}} \par}
\caption{\label{fig:dsde} Energy differential cross section for $^{11}$Be+$^{208}$Pb breakup at 69~MeV/u (top) and 520~MeV/u (bottom). The data from Refs.~\cite{Fuk04} and \cite{Pal03}  are compared with XCDCC calculations (solid lines) using S3 and S5 structure models and with EPM calculations (dashed lines) with S3 model. In the 520~MeV/u case, the estimated nuclear breakup is added to the EPM result. 
}
\end{center}
\end{figure}

Breakup energy distributions are shown in Fig.~\ref{fig:dsde}(a) for two angular ranges: $0^\circ<\theta_\text{c.m.}<1.3^\circ$, which is considered to be ``safe Coulomb'', and  
$0^\circ<\theta_\text{c.m.}<6^\circ$, where nuclear effects are relevant. The EPM calculation based on the $B(E1)$ distribution given by the model S3 reproduces well  both sets of experimental data for breakup energies around the peak ($\sim$0-1~MeV). However, it underestimates the cross sections  at higher excitation energies ($\sim$1-2~MeV). This could be due to limitations of the EPM dynamics, but also to limitations of the S3 structure model. To disentangle these effects, we compare the EPM result with an XCDCC calculation based on the same S3 structure model.
For the ``safe Coulomb" angular range, these XCDCC calculations are slightly larger than the EPM result, over all the energy range.  For the larger angular range, both calculations agree well at the peak, but the XCDCC calculation is significantly larger at higher excitation energies ($\e\sim$1-2 MeV), and agrees well with the data. 
Our conclusion is that there is no accurate ``safe Coulomb'' angular range, and that dynamical effects included in the XCDCC calculations are specially important for larger breakup energies and larger angles.  

For the experiment at 520~MeV/u, no angular distribution was extracted in \cite{Pal03} so we focus on the angle-integrated energy differential cross section,  presented in Fig.~\ref{fig:dsde}(b). The nuclear breakup contribution,  as estimated in~\cite{Pal03}, was added incoherently to the EPM
calculation. The resulting EPM distribution largely overestimates the data. By contrast, the XCDCC calculations, based on S3 and S5 models, have a better agreement with the data, with some overestimation of the former. 

\begin{figure}[tb]
\begin{center}
  {\centering \resizebox*{0.85\columnwidth}{!}{\includegraphics{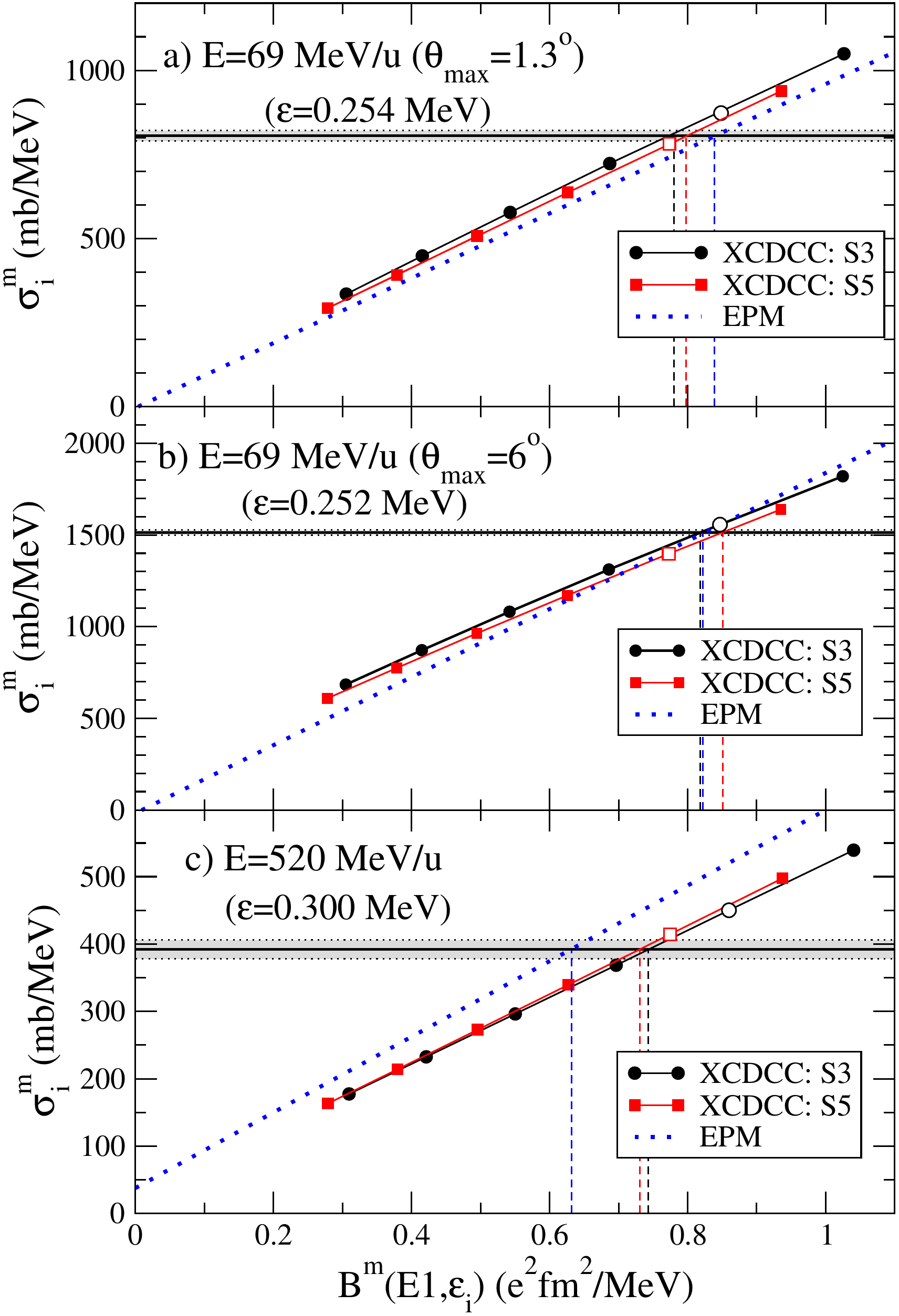}} \par}
\caption{\label{fig:sigma_vs_be} Relation between the $B(E1)$ values and associated cross sections computed with XCDCC (for S3 and S5 models) and with the EPM. The shaded area in each panel corresponds to the experimental cross sections, with the corresponding uncertainty. The vertical lines correspond to the extracted $B(E1)$ values. The symbols correspond to model calculations for different values of $\delta$, with the hollow ones corresponding to the $\delta=0$ cases. See text for details.} 
\end{center}
\end{figure}

\begin{figure}[tb]
\begin{center}
 {\centering \resizebox*{0.9\columnwidth}{!}{\includegraphics{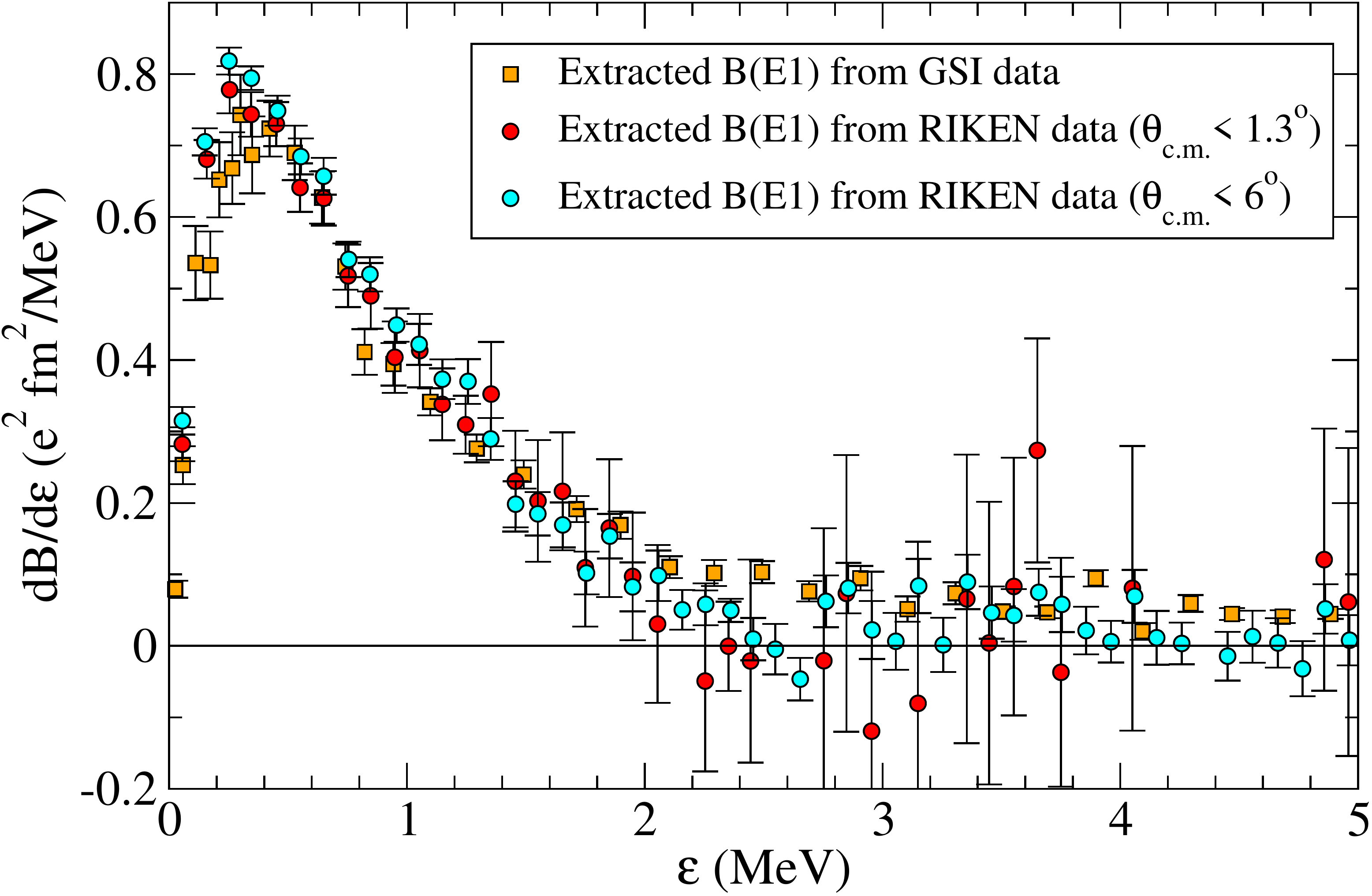}} \par}
\caption{\label{fig:BE1} $B(E1)$ distributions extracted from the experimental breakup data from Refs.~\cite{Pal03} and \cite{Fuk04} using the present method. 
}
\end{center}
\end{figure}

In Fig.~\ref{fig:sigma_vs_be} we illustrate the extraction of the $B(E1)$ from the experimental cross sections, with the method proposed in this work [Eq.~(\ref{Bexpnew})] and using the EPM [Eq.~(\ref{EPMexp})]. In the latter (dotted lines), the relation is strictly linear, and the slope is given by the dipole Coulomb excitation function, which is model independent [c.f.~Eq.~(\ref{EPM})]. 
The solid lines are the XCDCC cross sections for different initial $B(E1)$ distributions, obtained by scaling the dipole couplings by $\delta(\epsilon_i)$ factors ranging from $-0.4$ to $0.1$.
 It is seen that the relation between $\sigma_i$ and $B(E1,\varepsilon_i)$ is linear to a very good approximation, thus supporting Eq.~(\ref{quadraticnew}). 
However, the slope of the EPM lines differs  significantly from that of the XCDCC calculations, leading to markedly different extracted $B(E1)$ values. The slope of the XCDCC calculations, although model dependent, contains Coulomb-nuclear interference, as well as other dynamical effects which are absent in the EPM calculations. Note also that the $B(E1)$ values extracted using the EPM at the two different collision energies are significantly different, while those extracted from the XCDCC calculations are more compatible, as it will be shown in 
Fig.~\ref{fig:BE1}.

\begin{figure}[tb]
\begin{center}
  {\centering \resizebox*{0.85\columnwidth}{!}{\includegraphics{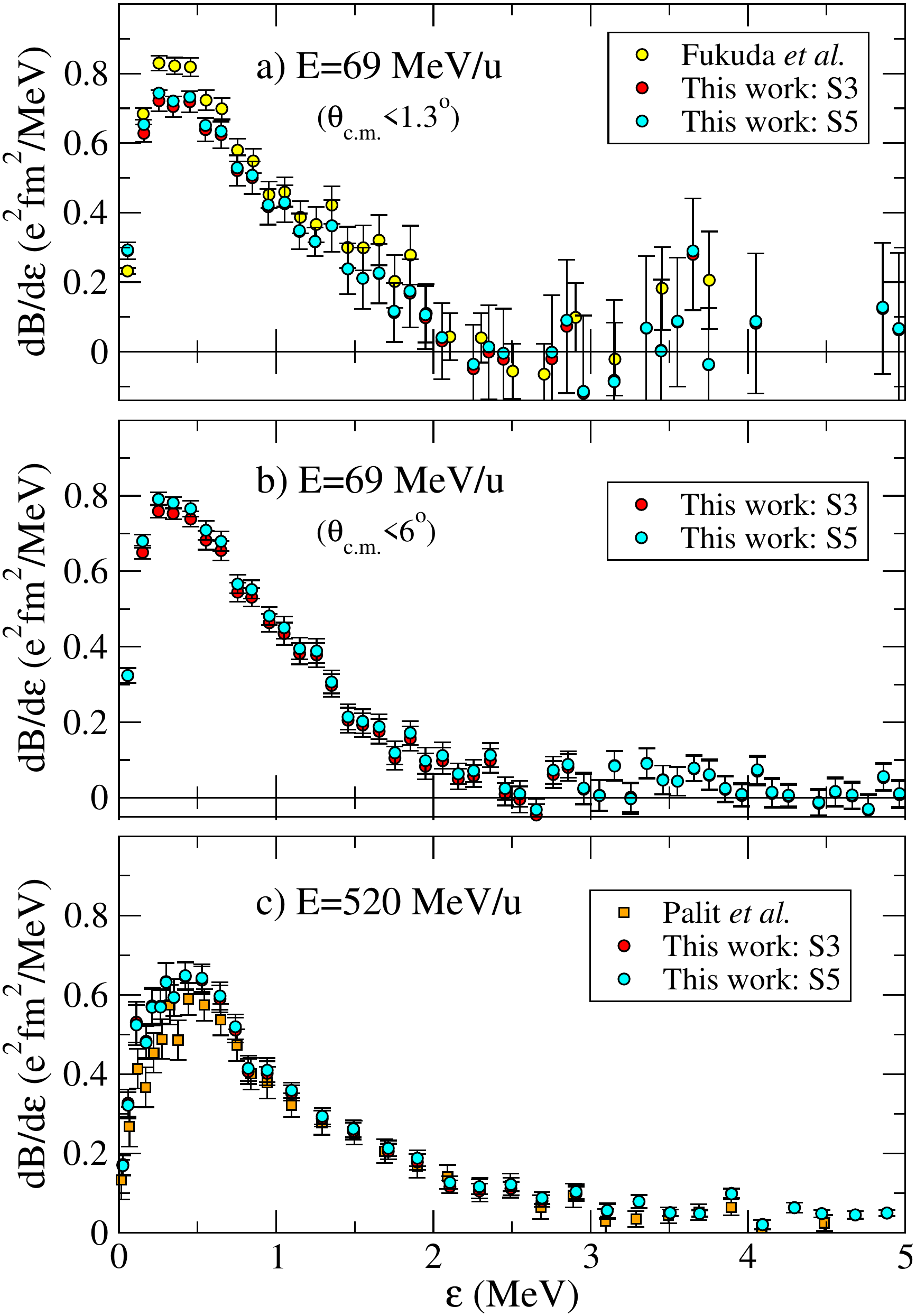}} \par}
\caption{\label{fig:BE1_folded} $B(E1)$ distributions extracted from the $^{11}$Be+$^{208}$Pb breakup data at 69~MeV/u and 520~MeV/u. The distributions reported in the original analyses \cite{Fuk04,Pal03} are compared with those extracted in the present work, starting with the structure models S3 and S5 described in the text. The  latter are convoluted with the experimental energy resolution.}
\end{center}
\end{figure}

In Fig.~\ref{fig:BE1_folded}, we compare the extracted $B(E1)$ distributions, using both S3 and S5 models,  with those obtained in the original analyses of the two considered experiments. 
For a meaningful comparison,  we have  convoluted the S3 and S5 $B(E1)$ distributions  with the resolution quoted  for each experiment, and then extracted a convoluted experimental $B(E1)$ distribution using Eq.~(\ref{Bexpnew}). For the RIKEN data, we show separately the results for the angular intervals  $\theta_\mathrm{c.m.} \leq 1.3^\circ$ (top panel) and  $\theta_\mathrm{c.m.} \leq 6^\circ$ (middle panel). In the former  we see that our derived $B(E1)$ agrees rather well with that from the original analysis for energies above 1~MeV, but  is somewhat lower at the peak, due to the dynamical effects discussed previously.
For the GSI data (bottom panel), our extracted $B(E1)$ agrees also very well with that of~\cite{Pal03} but are slightly higher at the peak. These two effects
go in the direction of making the results of the two experiments more compatible. It is also noticeable that the S3 and S5 models, while predicting rather different cross sections [c.f.~Figs.~\ref{fig:dsde} and \ref{fig:sigma_vs_be}], give rise to fully consistent $B(E1)$ distributions once they are corrected following the present procedure.  

Note that the error bars in the extracted $B(E1)$ include the experimental uncertainty of the cross sections only. We have performed a preliminary estimation of the systematic uncertainties introduced by the model dependence (comparing S3 and S5 models),  the choice of the nuclear potentials (using different prescriptions for fragment-target interactions) and  non-linearity in the relation between the $B(E1)$ and the cross section. These sources of systematic uncertainties are found to be similar or smaller than the experimental uncertainties. 
We expect to deepen the uncertainty analysis in future publications.

We have also extracted the experimental $B(E1)$ distribution using  Eq.~(\ref{Bexpnew}) starting with the original (i.e.\ unfolded) S3 model distribution of $B(E1)$. The results, which correspond to an unfolded experimental $B(E1)$ distribution, are shown in  Fig.~\ref{fig:BE1}.  For the high-energy data  at 520~MeV/u, our derived values are significant larger than those extracted in the original EPM analysis  \cite{Pal03}. This is partly due to the effect of the energy convolution. We also present the  $B(E1)$ distributions extracted  from the data at 69~MeV/u for the angular ranges $\theta_\text{c.m.}\leq 1.3^\circ$ and $\theta_\text{c.m.}\leq 6^\circ$.
 Note the relatively larger error bars for the smaller angular range, stemming from  the smaller cross sections for these interval. The $B(E1)$ extracted from the data up to $6^\circ$ are indeed affected by nuclear interaction, but  these effects are explicitly considered in our procedure. Notice the remarkable agreement of the three derived distributions. 
 Our extracted $B(E1)$ values from the two experiments turn out to be fully consistent and hence no discrepancy between the measured cross sections is apparent from our analysis. 
 
 The present results solve the long-standing controversy between these two measurements. Furthermore, it shows that a proper description of the reaction, including Coulomb and nuclear effects on an equal footing, is necessary for a meaningful extraction of structure information of the projectile. We consider that the present procedure for extracting $B(E1)$ distributions from Coulomb-dominated breakup cross section data can be applied to other exotic nuclei, which are currently being measured at radioactive beam facilities such as RIKEN, MSU and GSI-FAIR.

\bigskip
\section*{Acknowledgements}
We are grateful to Takashi Nakamura and Thomas Aumann for discussions on the RIKEN and GSI data analysed in this work, to Kazuyuki Ogata for useful feedback on the relativistic corrections and to T.~Furumoto and D.~Pang for providing us the double-folding g-matrix potentials employed in our calculations.   
This work has been partially supported by the Spanish Ministerio de Ciencia, Innovaci\'on y Universidades and FEDER funds under project FIS2017-88410-P and RTI2018-098117-B-C21 and by the European Union's Horizon 2020 research and innovation program under Grant Agreement No.\ 654002.

 \appendix

\section{Amplitude analysis} 
\label{apen:amplitudes}
In this section we justify the use of a correction factor determined from the differential cross sections, to obtain the experimental $B(E1)$ distribution.

Consider a model calculation, involving some nuclear and Coulomb couplings, which are considered to all orders. 
A general quantum mechanical treatment of Coulomb dissociation experiments leads to  cross sections which are given by an integral over the angular range and the energy resolution of the calculated double differential cross section
\begin{equation}
\label{sigma_res} 
 \sigma^0_i = \int dx \, R(\varepsilon_i,x)  \frac{d^2 \sigma^0}{d \Omega d \varepsilon},
\end{equation}
where $R(\varepsilon_i,x)$ represents the experimental angular and energy resolutions and  $x=(\Omega, \varepsilon)$ incorporates both the centre of mass angle and the energy of the break-up fragments. The double differential cross section, in turn, contains an average over the  ground state spin projection $N$, as well as a sum over the break-up states $M$ compatible with the energy $\varepsilon$, of the square of the transition amplitudes connecting them. The discrete index $M$ labels completely the break-up states, so it includes the core angular momentum, the halo neutron orbital angular momentum, the halo neutron total angular momentum, the total angular momentum of  the halo nucleus, and its spin projection.
\begin{equation}
\label{sigma_amp} 
\frac{d^2 \sigma^0}{d \Omega d \varepsilon} = \overline{\sum_{NM}} |A(N,M,x)|^2
\end{equation}
where $\overline{\sum_{NM}}$ indicates this sum over final states $M$ and average over the initial states $N$. 
 In Coulomb dominated breakup reactions, the amplitude $A(N,M,x)$, is dominated by a dipole Coulomb term which is proportional to the Coulomb dipole matrix element, so that  $A_C(N,M,x) = \langle N |{\cal M}(E1)|M,\varepsilon \rangle A_D(x)$,  but it will also have an extra term $A_n(N,M,x)$, containing the nuclear component, as well as other higher order Coulomb components, which we do not want to neglect. 
 Note that we do not have to make any assumption about the specific assumption about the expression $A_D(x)$. In particular, we no not need to make any semi-classical assumption, or neglect the effect of nuclear forces or other dynamical effects. We only assume that the general amplitude
 $A(N,M,x)$ has a dependence on the initial and final states $N,M$, which has a dominant term proportional to $ \langle N |{\cal M}(E1)|M,\varepsilon \rangle$, and an additional, smaller term $A_n(N,M,x)$, where the dependence on $N,M$ is different.
 The model will produce a $B(E1)$ distribution 
\begin{equation}
    {d B^0(E1, \varepsilon) \over d \varepsilon} = \overline{\sum_{NM}}  | \langle N|{\cal M}(E1)|M,\varepsilon \rangle|^2,
\end{equation}
which, particularized at the experimental energies, reads
\begin{equation}
     B^0(E1, \varepsilon_i) = \overline{\sum_{NM}}  |\langle N|{\cal M}(E1)|M,\varepsilon_i \rangle|^2,
\end{equation}
and the associated differential cross sections 
\begin{equation}
\label{amp2sigma}
\frac{d^2 \sigma^0}{d \Omega d \varepsilon} = \overline{\sum_{MN}} |\langle N|{\cal M}(E1)|M,\varepsilon \rangle A_D(x)  + A_n(N, M, x)|^2  .
\end{equation}

{Expression (\ref{amp2sigma})} indicates that the measured cross section is not proportional to the $B(E1)$ distribution, as assumed in the EPM. It also indicates that, owing to the presence of  interference terms,  it is not possible to estimate the nuclear effects as a nuclear cross section to be added to the pure Coulomb one. Despite of that, we will see how it is possible to obtain the $B(E1)$ distribution from these cross sections.

Consider now that we make arbitrary small changes in the model. This will result in small changes in the Coulomb dipole couplings connecting the ground state with the continuum states,  
that can be described by factors $(1+\delta(N,M,\varepsilon))$, where $\delta(N,M,\varepsilon)$ are arbitrary small numbers. The  nuclear couplings could in principle be also modified, producing small changes in $ A_n(N, M, x)$. However, as the  nuclear amplitides are already small compared to the Coulomb ones,  their small changes would be a second order effect, that can be neglected.  For this modified model, the $B(E1)$ distribution is given by
\begin{eqnarray}
\label{eq:delav}
    {d B^m(E1, \varepsilon) \over d \varepsilon} & \simeq & \overline{\sum_{MN}} (1+ 2 \delta(N,M,\varepsilon)) |\langle N|{\cal M}(E1)|M,\varepsilon \rangle|^2 \nonumber \\
    &=& (1+ 2 \delta(\varepsilon)) {d B^0(E1, \varepsilon) \over d \varepsilon} ,
\end{eqnarray}
where $\delta(\varepsilon)$ is a weighted average of the $\delta(NM,\varepsilon)$ values corresponding to the different dipole couplings between the ground state and the states with energy $\varepsilon$, i.e.,
\begin{equation}
\label{delta} 
\delta(\varepsilon) = {\overline{\sum_{NM}} \delta(N,M,\varepsilon) |\langle N|{\cal M}(E1)|M,\varepsilon \rangle|^2 \over  \overline{\sum_{NM}} |\langle N|{\cal M}(E1)|M,\varepsilon \rangle|^2} .
\end{equation}

This expression can be particularized at the experimental energies $\varepsilon_i$, leading to
\begin{equation}
 B^m(E1, \varepsilon_i)  \simeq (1+ 2 \delta(\varepsilon_i)) B^0(E1, \varepsilon_i) .
 \label{Bmodified}
\end{equation}

Let us  now consider the effect on the cross sections. The modified differential cross sections are
\begin{align}
\sigma^m_i =& \int dx R(\varepsilon_i,x) 
\overline{\sum_{MN}} |(1 + \delta(N,M,\varepsilon)) A_C(N,M,x)  + A_n(N, M, x)|^2 .
\end{align}

This expanded to the lowest order in the small parameters $\delta(N,M,\varepsilon)$. Also, considering that the energy range for $R(\varepsilon_i,x)$ is sufficiently narrow, compared to the energy dependence of the electric matrix elements, can get $\delta(N,M,\varepsilon)$ and $\langle N|{\cal M}(E1)|M,\varepsilon \rangle$ out of the integral, and evaluate them at the nominal energy $\varepsilon_i$. Note, however, that we do not need to make any assumption about the energy dependence of the amplitude $A_D(x)$.
\begin{align}
\sigma^m_i & \simeq  \sigma^0_i + \overline{\sum_{MN}} \delta(N,M,\varepsilon_i) |\langle N|{\cal M}(E1)|M,\varepsilon_i \rangle|^2  
 \nonumber \\
&\times \int dx R(\varepsilon_i,x) 
 \left(|A_D(x)|^2  +  {A_D(x)A^*_n(N, M,X) + cc \over  \langle N|{\cal M}(E1)|M,\varepsilon_i \rangle} \right) . \label{sigma_modified} 
\end{align}
In a Coulomb-dominated reaction, the dipole amplitude $|A_D(x)|^2$ dominates over the Coulomb-nuclear interference term, and hence the term in parenthesis is approximately independent of the final dipole state. This justifies replacing  $\delta(N,M,\varepsilon_i)$ by the weighted average $\delta(\varepsilon_i)$ given by Eq.~(\ref{delta}). It also justifies neglecting any small correction of the nuclear amplitudes. Thus we get
\begin{equation}
   \sigma^m_i  \simeq  \sigma^0_i + \delta(\varepsilon_i) \sigma'_i , \label{sigmadelta}
\end{equation}
where
\begin{align}
\sigma'_i & =   \overline{\sum_{MN}} |\langle N|{\cal M}(E1)|M,\varepsilon_i \rangle|^2   \nonumber \\
& \times \int dx \,  R(\varepsilon_i,x)  \left(|A_D(x)|^2  +  {A_D(x)A^*_n(N,M,x) + cc \over  \langle N|{\cal M}(E1)|M,\varepsilon_i \rangle} \right) .
\label{sigma_modified_2} 
\end{align}
The practical calculation of  $\sigma'_i$  can be done evaluating the cross sections $\sigma_i(\delta)$ at the experimental energies $\varepsilon_i$ from model calculations where all the dipole couplings have been renormalized  by factors $(1+\delta)$,  using small values of $\delta$, such as $\delta = \pm 0.1$. 
\begin{equation}
 \label{sigmap}
\sigma'_i = {1 \over 0.2} \left( \sigma_i(\delta=0.1) - \sigma_i (\delta=-0.1) \right).
\end{equation}

\section*{References}
\bibliography{xcdcc}

\begin{thebibliography}{10}
\expandafter\ifx\csname url\endcsname\relax
  \def\url#1{\texttt{#1}}\fi
\expandafter\ifx\csname urlprefix\endcsname\relax\def\urlprefix{URL }\fi
\expandafter\ifx\csname href\endcsname\relax
  \def\href#1#2{#2} \def\path#1{#1}\fi

\bibitem{AW75}
K.~Alder, A.~Winther, Electromagnetic excitation: theory of Coulomb excitation
  with heavy ions, North-Holland Pub. Co., 1975.

\bibitem{Ber88}
C.~A. Bertulani, G.~Baur, {Electromagnetic processes in relativistic heavy ion
  collisions}, Phys. Rep. 163 (1988) 299.
\newblock \href {http://dx.doi.org/10.1016/0370-1573(88)90142-1}
  {\path{doi:10.1016/0370-1573(88)90142-1}}.

\bibitem{Gla98}
T.~Glasmacher, Coulomb excitation at intermediate energies, Annu. Rev. Nucl.
  Part. Sci. 48 (1998) 1.
\newblock \href {http://dx.doi.org/10.1146/annurev.nucl.48.1.1}
  {\path{doi:10.1146/annurev.nucl.48.1.1}}.

\bibitem{Gad03}
A.~Gade, D.~Bazin, C.~M. Campbell, J.~A. Church, D.~C. Dinca, J.~Enders,
  T.~Glasmacher, Z.~Hu, K.~W. Kemper, W.~F. Mueller, H.~Olliver, B.~C. Perry,
  L.~A. Riley, B.~T. Roeder, B.~M. Sherrill, J.~R. Terry, Detailed experimental
  study on intermediate-energy coulomb excitation of ${}^{46}\mathrm{Ar}$,
  Phys. Rev. C 68 (2003) 014302.
\newblock \href {http://dx.doi.org/10.1103/PhysRevC.68.014302}
  {\path{doi:10.1103/PhysRevC.68.014302}}.

\bibitem{Esb95}
H.~Esbensen, G.~Bertsch, C.~Bertulani, {Higher-order dynamical effects in
  Coulomb dissociation}, Nucl. Phys. A 581 (1995) 107.
\newblock \href {http://dx.doi.org/10.1016/0375-9474(94)00423-K}
  {\path{doi:10.1016/0375-9474(94)00423-K}}.

\bibitem{Esb02}
H.~Esbensen, G.~F. Bertsch, {Higher-order effects in the two-body breakup of
  $^{17}$F}, Nucl. Phys. A 706 (2002) 383.
\newblock \href {http://dx.doi.org/10.1016/S0375-9474(02)00869-2}
  {\path{doi:10.1016/S0375-9474(02)00869-2}}.

\bibitem{Pes17}
V.~Pesudo, et~al., Scattering of the halo nucleus $^{11}\mathrm{Be}$ on
  $^{197}\mathrm{Au}$ at energies around the coulomb barrier, Phys. Rev. Lett.
  118 (2017) 152502.
\newblock \href {http://dx.doi.org/10.1103/PhysRevLett.118.152502}
  {\path{doi:10.1103/PhysRevLett.118.152502}}.

\bibitem{Pal03}
R.~Palit, et~al., Exclusive measurement of breakup reactions with the
  one-neutron halo nucleus ${}^{11}\mathrm{Be}$, Phys. Rev. C 68 (2003) 034318.
\newblock \href {http://dx.doi.org/10.1103/PhysRevC.68.034318}
  {\path{doi:10.1103/PhysRevC.68.034318}}.

\bibitem{Fuk04}
N.~Fukuda, et~al., {Coulomb and nuclear breakup of a halo nucleus $^{11}$Be},
  Phys. Rev. C 70 (2004) 054606.
\newblock \href {http://dx.doi.org/10.1103/PhysRevC.70.054606}
  {\path{doi:10.1103/PhysRevC.70.054606}}.

\bibitem{Sum06}
N.~C. {Summers}, F.~M. {Nunes}, I.~J. {Thompson}, {Extended continuum
  discretized coupled channels method: Core excitation in the breakup of exotic
  nuclei}, Phys. Rev. C 74 (2006) 014606.
\newblock \href {http://dx.doi.org/10.1103/PhysRevC.74.014606}
  {\path{doi:10.1103/PhysRevC.74.014606}}.

\bibitem{Die14}
R.~{de Diego}, J.~M. {Arias}, J.~A. {Lay}, A.~M. {Moro}, Continuum-discretized
  coupled-channels calculations with core excitation, Phys. Rev. C 89 (2014)
  064609.
\newblock \href {http://dx.doi.org/10.1103/PhysRevC.89.064609}
  {\path{doi:10.1103/PhysRevC.89.064609}}.

\bibitem{Cal16}
A.~Calci, P.~Navr\'atil, R.~Roth, J.~Dohet-Eraly, S.~Quaglioni, G.~Hupin, Can
  ab initio theory explain the phenomenon of parity inversion in
  $^{11}\mathrm{Be}$?, Phys. Rev. Lett. 117 (2016) 242501.
\newblock \href {http://dx.doi.org/10.1103/PhysRevLett.117.242501}
  {\path{doi:10.1103/PhysRevLett.117.242501}}.

\bibitem{Mos19}
L.~Moschini, P.~Capel, {Reliable extraction of the $dB(E1)/dE$ for $^{11}$Be
  from its breakup at 520 MeV/nucleon}, Phys. Lett. B 790 (2019) 367.
\newblock \href {http://dx.doi.org/10.1016/j.physletb.2019.01.041}
  {\path{doi:10.1016/j.physletb.2019.01.041}}.

\bibitem{Sum07}
N.~Summers, S.~Pain, N.~Orr, W.~Catford, J.~Angélique, N.~Ashwood, V.~Bouchat,
  N.~Clarke, N.~Curtis, M.~Freer, B.~Fulton, F.~Hanappe, M.~Labiche,
  J.~Lecouey, R.~Lemmon, D.~Mahboub, A.~Ninane, G.~Normand, F.~Nunes, N.~Soić,
  L.~Stuttge, C.~Timis, I.~Thompson, J.~Winfield, V.~Ziman, {$B(E1)$ strengths
  from Coulomb excitation of $^{11}$Be}, Phys. Lett. B 650 (2007) 124.
\newblock \href
  {http://dx.doi.org/https://doi.org/10.1016/j.physletb.2007.05.003}
  {\path{doi:https://doi.org/10.1016/j.physletb.2007.05.003}}.

\bibitem{Ger83}
H.~V. von Geramb, Microscopic optical potentials, AIP Conf. Proc 97 (1983) 44.
\newblock \href {http://dx.doi.org/10.1063/1.33973}
  {\path{doi:10.1063/1.33973}}.

\bibitem{Rik84}
L.~Rikus, K.~Nakano, H.~V. Von~Geramb, {Microscopic analysis of elastic and
  inelastic proton scattering from $^{12}$C}, Nucl. Phys. A 414 (1984) 413.
\newblock \href {http://dx.doi.org/10.1016/0375-9474(84)90611-0}
  {\path{doi:10.1016/0375-9474(84)90611-0}}.

\bibitem{KD03}
A.~Koning, J.~Delaroche, {Local and global nucleon optical models from 1 keV to
  200 MeV}, Nucl. Phys. A 713 (2003) 231.
\newblock \href {http://dx.doi.org/10.1016/S0375-9474(02)01321-0}
  {\path{doi:10.1016/S0375-9474(02)01321-0}}.

\bibitem{Bau98}
E.~Bauge, J.~P. Delaroche, M.~Girod, {Semimicroscopic nucleon-nucleus spherical
  optical model for nuclei with $A>\sim$40 at energies up to 200 MeV}, Phys.
  Rev. C 58 (1998) 1118.
\newblock \href {http://dx.doi.org/10.1103/PhysRevC.58.1118}
  {\path{doi:10.1103/PhysRevC.58.1118}}.

\bibitem{Bau01}
E.~Bauge, J.~P. Delaroche, M.~Girod, {Lane-consistent, semimicroscopic
  nucleon-nucleus optical model}, Phys. Rev. C 63 (2001) 024607.
\newblock \href {http://dx.doi.org/10.1103/PhysRevC.63.024607}
  {\path{doi:10.1103/PhysRevC.63.024607}}.

\bibitem{Pang-private}
D.~Y. Pang, {Private Communication}.

\bibitem{Xu13}
Y.~P. Xu, D.~Y. Pang, Toward a systematic nucleus-nucleus potential for
  peripheral collisions, Phys. Rev. C 87 (2013) 044605.
\newblock \href {http://dx.doi.org/10.1103/PhysRevC.87.044605}
  {\path{doi:10.1103/PhysRevC.87.044605}}.

\bibitem{Furu-private}
T.~Furumoto, {Private Communication}.

\bibitem{Fur12}
T.~Furumoto, W.~Horiuchi, M.~Takashina, Y.~Yamamoto, Y.~Sakuragi, {Global
  optical potential for nucleus-nucleus systems from 50 MeV/u to 400 MeV/u},
  Phys. Rev. C 85 (2012) 044607.
\newblock \href {http://dx.doi.org/10.1103/PhysRevC.85.044607}
  {\path{doi:10.1103/PhysRevC.85.044607}}.

\bibitem{Mor15}
A.~M. Moro, Three-body model for the analysis of quasifree scattering reactions
  in inverse kinematics, Phys. Rev. C 92 (2015) 044605.
\newblock \href {http://dx.doi.org/10.1103/PhysRevC.92.044605}
  {\path{doi:10.1103/PhysRevC.92.044605}}.

\end{thebibliography}

\end{document}